\documentclass[aps,prd,twocolumn,superscriptaddress]{revtex4-1}

\usepackage{longtable}
\usepackage{grffile} 
\usepackage{graphics}
\usepackage{graphicx} 
\usepackage{amsmath}    
\usepackage{hyperref}   
\usepackage{subfigure}  
\usepackage{epsfig}
\usepackage{epstopdf}
\usepackage{ulem}
\usepackage{color}
\usepackage{booktabs}

\begin{document}

\title{Strangeness production in $p$Pb collisions at $\sqrt{s_{NN}} = 5.02$ TeV in the PACIAE model}

\author{Wei Lei}
\affiliation{Key Laboratory of Quark and Lepton Physics (MOE) and \break Institute
	of Particle Physics, Central China Normal University,\break Wuhan 430079, China}

\author{Qiang Wang}
\affiliation{Key Laboratory of Quark and Lepton Physics (MOE) and \break Institute
	of Particle Physics, Central China Normal University,\break Wuhan 430079, China}

\author{An-Ke Lei}\email{ankelei@mails.ccnu.edu.cn}
\affiliation{Key Laboratory of Quark and Lepton Physics (MOE) and \break Institute
	of Particle Physics, Central China Normal University,\break Wuhan 430079, China}
\affiliation{School of Physics and Electronic Science, Guizhou Normal University, Guiyang, 550025, China}

\author{Zhi-Lei She}
\affiliation{School of Mathematics and Statistics, Wuhan Textile University,  Wuhan
 430200, China}

\author{Jia-Xin Gao}
\affiliation{Key Laboratory of Quark and Lepton Physics (MOE) and \break Institute
	of Particle Physics, Central China Normal University,\break Wuhan 430079, China}

\author
{Dai-Mei Zhou}\email{zhoudm@mail.ccnu.edu.cn, corresponding author.}
\affiliation{Key Laboratory of Quark and Lepton Physics (MOE) and \break Institute of Particle Physics, Central China Normal University,\break Wuhan 430079, China}

\author{Yu-Liang Yan}\email{yuliang86@yeah.net}
\affiliation{China Institute of Atomic Energy, P. O. Box 275 (18), Beijing 102413, China}

\author{Hua Zheng}
\affiliation{School of Physics and Information Technology, Shaanxi Normal University,  Xi’an 710119, China}

\author{Wen-Chao Zhang}
\affiliation{School of Physics and Information Technology, Shaanxi Normal University,  Xi’an 710119, China}

\author{Ben-Hao Sa}\email{sabhliuym35@qq.com}
\affiliation{China Institute of Atomic Energy, P. O. Box 275 (18), Beijing 102413, China}
\affiliation{Key Laboratory of Quark and Lepton Physics (MOE) and \break Institute of Particle Physics, Central China Normal University,\break Wuhan 430079, China}

\date{\today}

\begin{abstract}
We present a study of the single- and multi-strange particle productions in $p$Pb collisions at $\sqrt{s_{NN}} = 5.02$ TeV using the PACIAE 4.0 model. The effects of the color reconnection (CR), rope hadronization (RH) as well as the partonic and hadronic rescatterings (PRS and HRS) on their productions are investigated, respectively. There are two kinds of CR considered: the multiparton interactions-based CR (MPI-CR) and the quantum chromodynamics-based CR (QCD-CR). 
The four mechanisms studied have little effect on kaon production.
However, the QCD-CR scheme incorporating the junction topology gives an enhanced baryon production, but it is still not enough to describe the integrated yields of strange baryons. The combination of QCD-CR and RH scheme together with the flavor ropes (FR) and the string shoving (SS), i.e., QCD-CR+FR+SS, provides a good description of the production of strange baryons. PRS has a weak effect but HRS exerts a promoting effect on the production of multi-strange baryons ($\Xi$ and $\Omega$). 
\end{abstract}

\maketitle

\section{Introduction}
The novel state of quark-gluon plasma (QGP, composed of the quarks and gluons) was originally conjectured as a transient state of strongly interacting matter that existed just microseconds after the Big Bang~\cite{ALICE:2022wpn}. In either the Relativistic Heavy Ion Collider (RHIC) experiments or the Large
Hadron Collider (LHC) experiments, the QGP-related information is only
hidden in the Final Hadronic State (FHS). Several observables are considered as evidences of QGP formation, including the mass hierarchy of the elliptic flow ($v_{2}$) at low transverse momentum
($p_{T}$)~\cite{ALICE:2013snk,CMS:2018loe}, the energy loss (jet quenching)~\cite{Li:2025ugv}, the polarization effect of $\Lambda$ hyperons~\cite{Lei:2021mvp}, 
and the strangeness enhancement~\cite{Rafelski:1982pu,Koch:1986ud}, etc. When QGP is formed, the chemically equilibrated system generates copious 
strange quark pairs through thermal gluon fusion processes enhancing the 
formation of multi-strange hadrons 
especially ~\cite{Rafelski:1982pu,Koch:1986ud}. Therefore, the strangeness enhancement, particularly that of multi-strange baryons, is widely considered as a distinctive signature of QGP
formation~\cite{Andersen:1998vu,ALICE:2013xmt,STAR:2003jis}.

The proton-proton ($pp$) and proton-nucleus ($p$A) collisions serve as
baselines for investigating QGP formation in heavy-ion collisions at the RHIC and LHC energies. It is supported by the ALICE observations of a significant enhancement in the strange hadron to pion yield ratio for the high-multiplicity $pp$ and $p$A collisions at LHC 
energies~\cite{ALICE:2016fzo,ALICE:2019avo,ALICE:2015mpp,ALICE:2013wgn}, which is similar to those observed in Pb-Pb~\cite{ALICE:2013mez,ALICE:2013cdo,ALICE:2013xmt} collisions at the same energies.

The universal characteristics of the strangeness-to-pion ratio as a function of multiplicity are interpreted through various models. The THERMUS model explains the reduced strange hadron yields in $pp$ collisions via the canonical strangeness conservation~\cite{Vislavicius:2016rwi}. The DIPSY 
model~\cite{Bierlich:2014xba} employs Mueller’s dipole formulation to describe strangeness production in a QCD-inspired approach, and accurately reproduces the yields of single-strange hadrons ($K_{s}^{0}$ and $\Lambda $). However, it systematically underestimates the multi-strange baryons ($\Xi$ and $\Omega$). 
The model attributes the enhanced strangeness production to the color ropes, which arises from interacting gluonic strings and increases the effective string tension~\cite{Biro:1984cf,Andersson:1991er}. The EPOS 
model~\cite{Pierog:2013ria} incorporates partial QGP formation in $pp$ 
collisions through a core-corona treatment of particle interactions. In the Monte Carlo event generator PACIAE 2.0~\cite{Sa:2011ye}, the mechanisms of single-string dynamics and multiple-string interactions, have successfully described the strange enhancement in the $pp$~\cite{Zheng:2018yxq} and the Pb-Pb collisions~\cite{Zhou:2020kwp}. Although current models successfully describe several aspects of the data, the underlying physics driving strangeness enhancement is not yet fully understood. This highlights the need for additional developments to obtain a complete microscopic description of strangeness production in small collision systems ($pp$ and $p$A) and to evaluate its possible role in heavy-ion collisions.

Recently, a combined mechanism of the color reconnection 
(CR)~\cite{OrtizVelasquez:2013ofg} and the rope hadronization 
(RH)~\cite{Bierlich:2017sxk,Bierlich:2014xba} has been used to explore the 
strangeness enhancement in $pp$ collisions at $\sqrt{s_{NN}} = 13$ TeV ~\cite{Hushnud:2023mgy}. This approach offers a novel perspective for understanding the microscopic dynamics of high-multiplicity $pp$ collision events observed at the LHC within the string fragmentation framework. The enhancement of strangeness was also studied in $pp$ collisions using non-perturbative mechanisms of RH, including string shoving (SS) and flavor ropes (FR)~\cite{Nayak:2018xip,Chakraborty:2021nde}. In the present study, we incorporate these aforementioned mechanisms into $p$Pb collisions at 
$\sqrt{s_{NN}}$ = 5.02 TeV using the microscopic transport model 
PACIAE 4.0~\cite{Lei:2024kam} and investigate the production of strange and multi-strange hadrons.

The paper is organized as follows. In Sec.~\ref{sec:model}, we present the PACIAE 4.0 model setup focusing on its implementation of CR and RH mechanisms, along with the partonic and hadronic rescatterings 
(PRS and HRS) processes. The multiplicity dependent $\langle dN/dy \rangle$, and the strange hadron-to-pion ratios are presented 
in Sec.~\ref{sec:results}. Finally, in Sec.~\ref{sec:conclusions}, we 
summarize the important results of the present work.

\begin{figure}[hbt!]
   \centering
        \subfigure{	\includegraphics[width=0.45\textwidth]{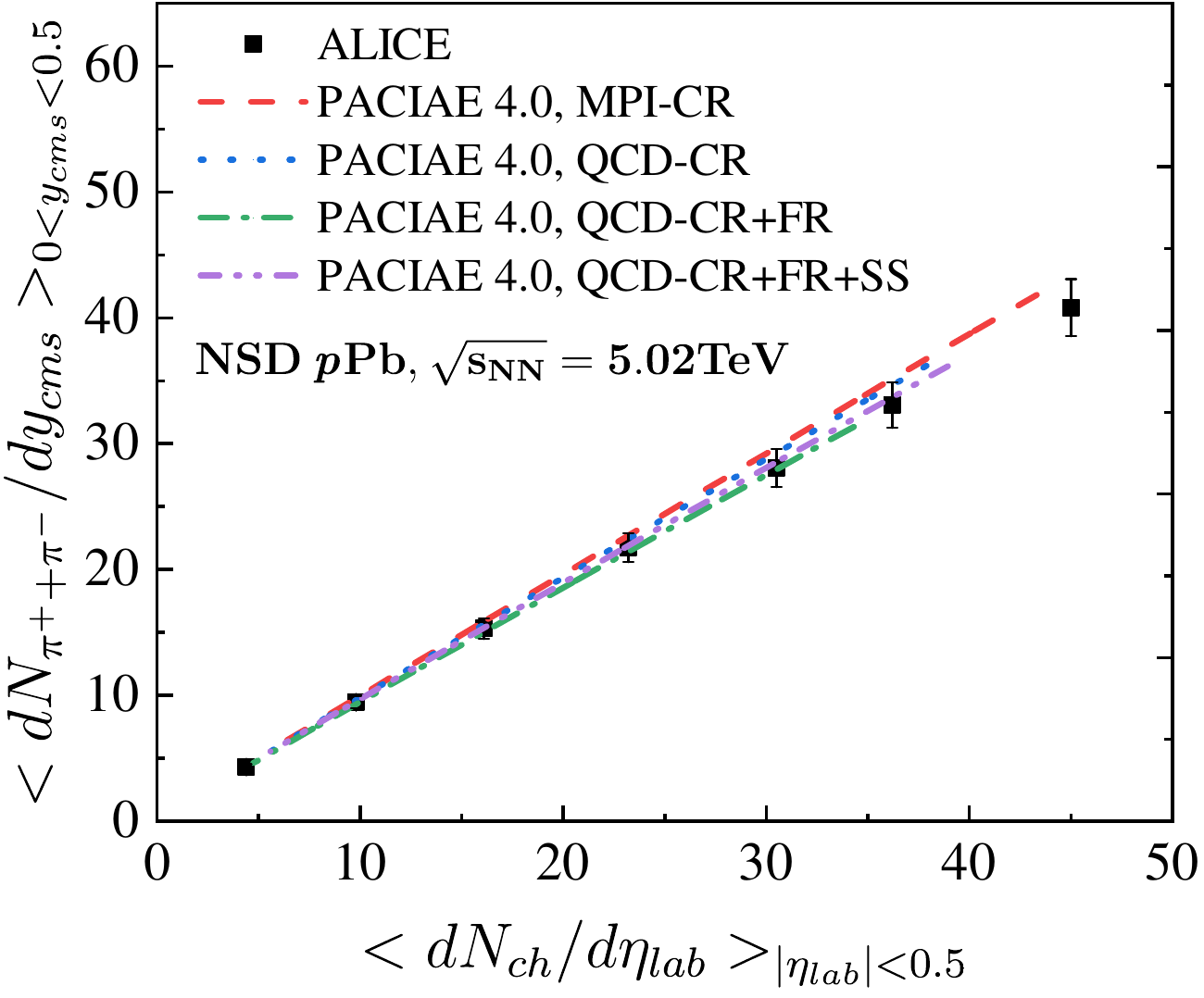}
		\label{fig1:a}
	}               
	\caption{ (Color online) Multiplicity dependent $\langle dN/dy_{cms} \rangle$ of $\pi$, varying with the $\langle dN_{ch}/d\eta_{lab} \rangle_{|\eta_{lab}|<0.5}$ in NSD $p$Pb collisions at $\sqrt{s_{NN}}=5.02$ TeV.  Four different CR and RH schemes are shown:
MPI-CR, red dashed line; QCD-CR, blue dotted line; QCD-CR+FR, green dash-dotted line; QCD-CR+FR+SS, purple dash-dot-dotted line. The experimental data are taken from Ref.~\cite{ALICE:2013wgn}. }
	\label{fig:PACIAE-pi+ks0+p}
\end{figure}

\section{THE SIMULATION FRAMEWORK}
\label{sec:model}

\subsection{PACIAE 4.0 Model}

PACIAE 4.0~\cite{Lei:2024kam} is a new generation of the Monte Carlo event 
generator PACIAE~\cite{Sa:2011ye,Lei:2023srp} for simulating high-energy 
lepton-lepton, lepton-hadron, lepton-nucleus, hadron-hadron, hadron-nucleus, and nucleus-nucleus collisions. It comprises four distinct stages: (1) initial partonic state, (2) partonic rescatterings, (3) hadronization, and (4) hadronic rescatterings. The PACIAE 4.0 is based on the PYTHIA 8~\cite{Sjostrand:2008za, Bierlich:2022pfr,Bierlich:2016smv,Bierlich:2018xfw}. The initial partonic 
state is generated by temporarily deactivating the Lund string fragmentation in PYTHIA. This initial state then undergoes elastic $2 \rightarrow 2$ partonic rescatterings with the leading-order perturbative QCD (LO-pQCD) cross sections~\cite{Combridge:1977dm,Combridge:1978kx}. After hadronization, the colored partons are converted to color-neutral hadrons. Finally, the produced hadrons experience $2 \rightarrow 2$ hadronic rescatterings with experimentally and/or empirically determined 
cross sections~\cite{Koch:1986ud,Schopper:1988hwx,Bierlich:2022pfr}.



\subsection{Mechanisms of CR and RH}

In nucleon-nucleon ($NN$) collisions, the multiparton interactions (MPI) 
stemming from the composite property of nucleons lead to an increase number of color charges in the system. This results in the rearrangement and reconnection of color strings minimizing the total string length, a 
phenomenon known as color reconnection (CR)~\cite{OrtizVelasquez:2013ofg}. 
This mechanism is promising in the explanation of heavy-ion-like phenomena 
observed in high-multiplicity $pp$ 
collisions~\cite{OrtizVelasquez:2013ofg,Bierlich:2015rha,Hushnud:2023mgy}. In this work, two distinct CR models are implemented to address different aspects of string dynamics in the high-energy collisions. One is the default MPI-based CR (MPI-CR) which reduces the total string length by a complete merge of the partons of separate MPI systems with the merging possibility related to the $p_T$ scale~\cite{Bierlich:2022pfr}. The other is QCD-based CR (QCD-CR), which determines the possibility of reconnection through the color symmetry principles of QCD~\cite{Christiansen:2015yqa}. In QCD-CR, a fraction of the string dipoles reconnect in junction configurations, which can enhance baryon production~\cite{Christiansen:2015yqa}.

PACIAE 4.0 provides two hadronization models: The Lund string fragmentation regime and the phenomenological coalescence model. In this work, the former hadronization mechanism is chosen~\cite{Andersson:1983ia}, which describes the formation and fragmentation of a color-confined string between interacting partons. A massless string is created by the color field between two separating partons, leading to linear confinement. As the partons move apart, 
the increasing string potential energy causes the string to break. The 
quark-antiquark or diquark-antidiquark pairs will be excited from the QCD 
vacuum via the quantum tunneling mechanism. Then the string fragments into 
hadrons. The rope hadronization (RH) framework~\cite{Bierlich:2014xba} represents an advanced extension of the conventional Lund string fragmentation regime. It incorporates two distinct but complementary string interaction mechanisms: The flavor ropes (FR) formation and the string shoving (SS) dynamics. The FR mechanism allows the formation of ropes between overlapping strings in dense environments. 
The string ropes can be hadronized by larger effective string 
tension~\cite{Bierlich:2016vgw}, which increases the tunneling probabilities of strange quark-antiquark and diquark-antidiquark pairs, enhancing the production of strange hadrons~\cite{Biro:1984cf,Bierlich:2014xba}. The SS 
mechanism~\cite{Bierlich:2017vhg,Bierlich:2020naj} achieves mutual pushing 
effects between adjacent color strings through an interaction potential 
analogous to the color superconductivity~\cite{Bierlich:2022pfr}. It is 
shown that SS would occur before FR~\cite{Bierlich:2024odg}. Because the SS spatially pushes strings outward, reducing the string-overlap, it will diminish the FR effect. The combined FR+SS will provide slightly smaller string tension comparing to the FR.

\begin{figure*}[hbt!]
	\centering

        \subfigure{
		\includegraphics[width=0.45\textwidth]{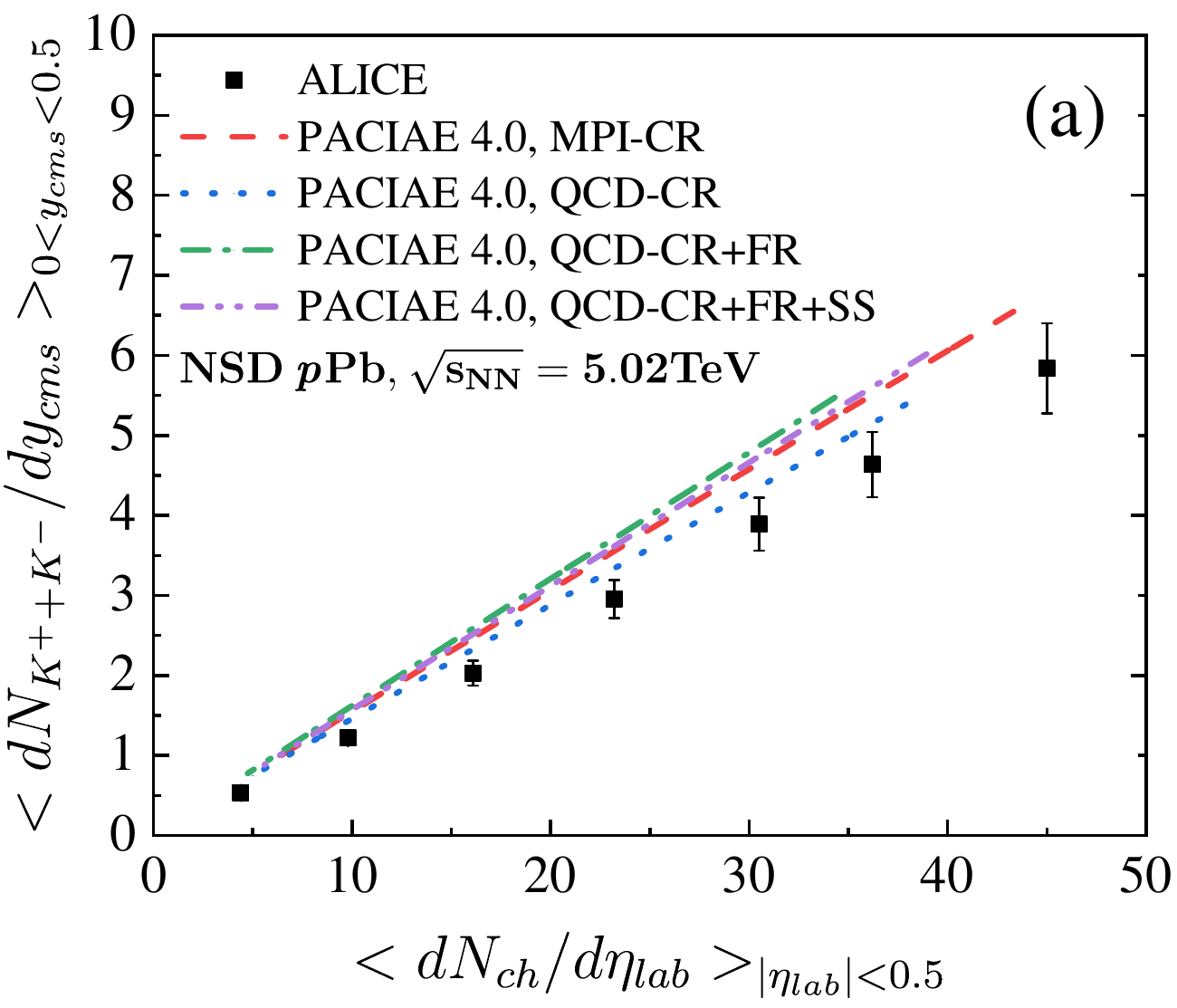}
		\label{fig2:a}
	}
        \subfigure{
		\includegraphics[width=0.45\textwidth]{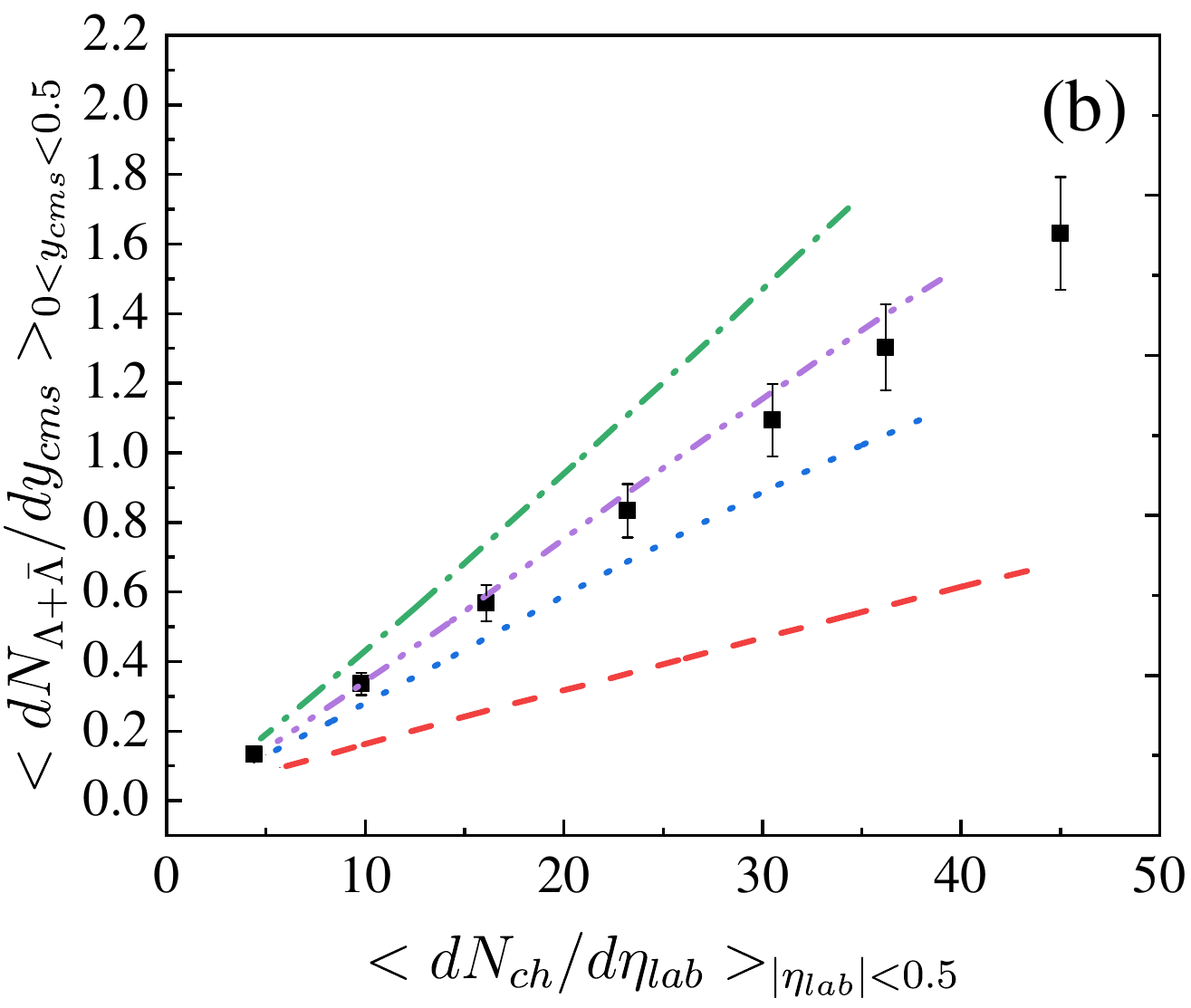}
		\label{fig2:b}
	}
        \subfigure{	\includegraphics[width=0.45\textwidth]{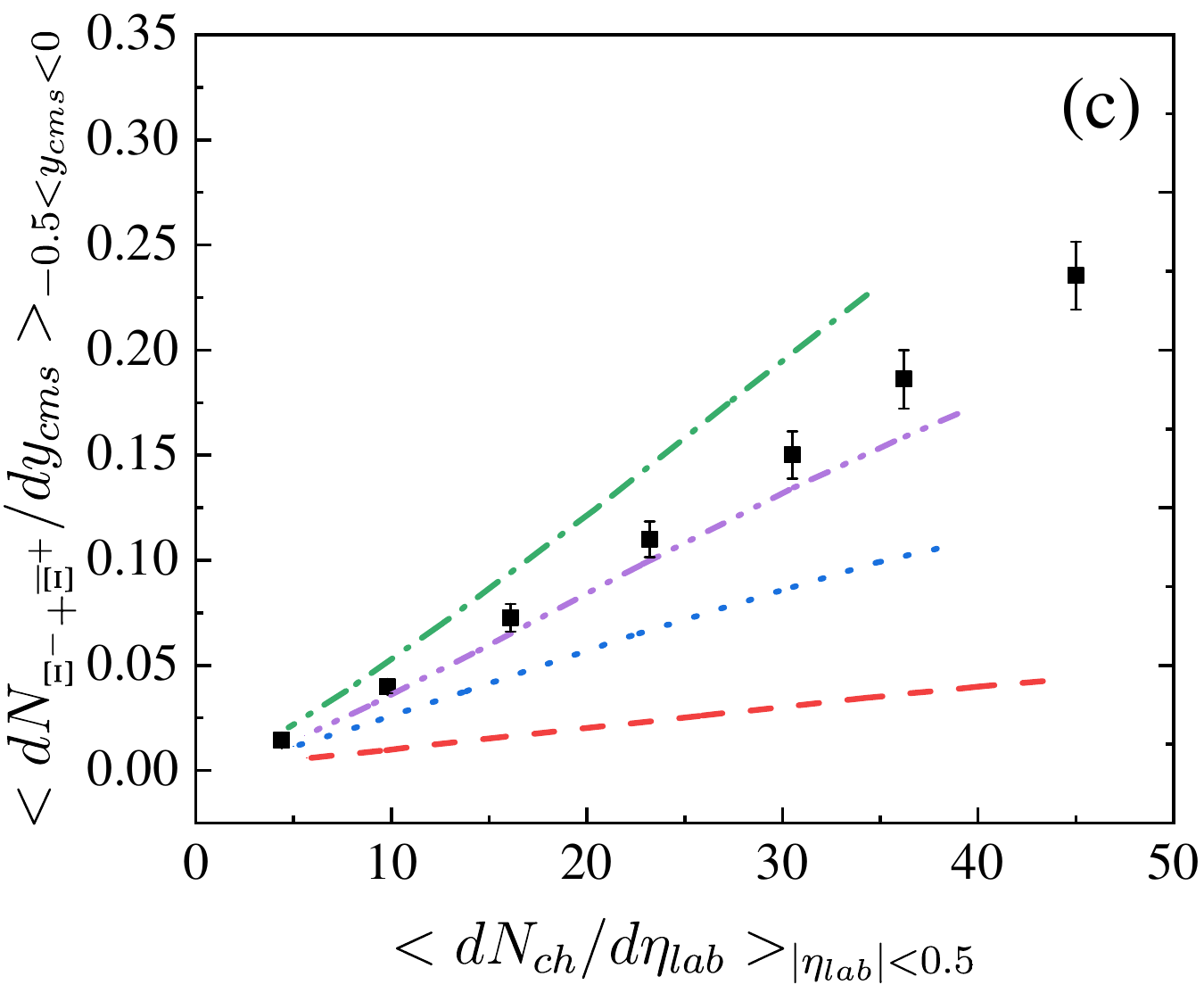}
		\label{fig2:c}
        }
        \subfigure{
		\includegraphics[width=0.45\textwidth]{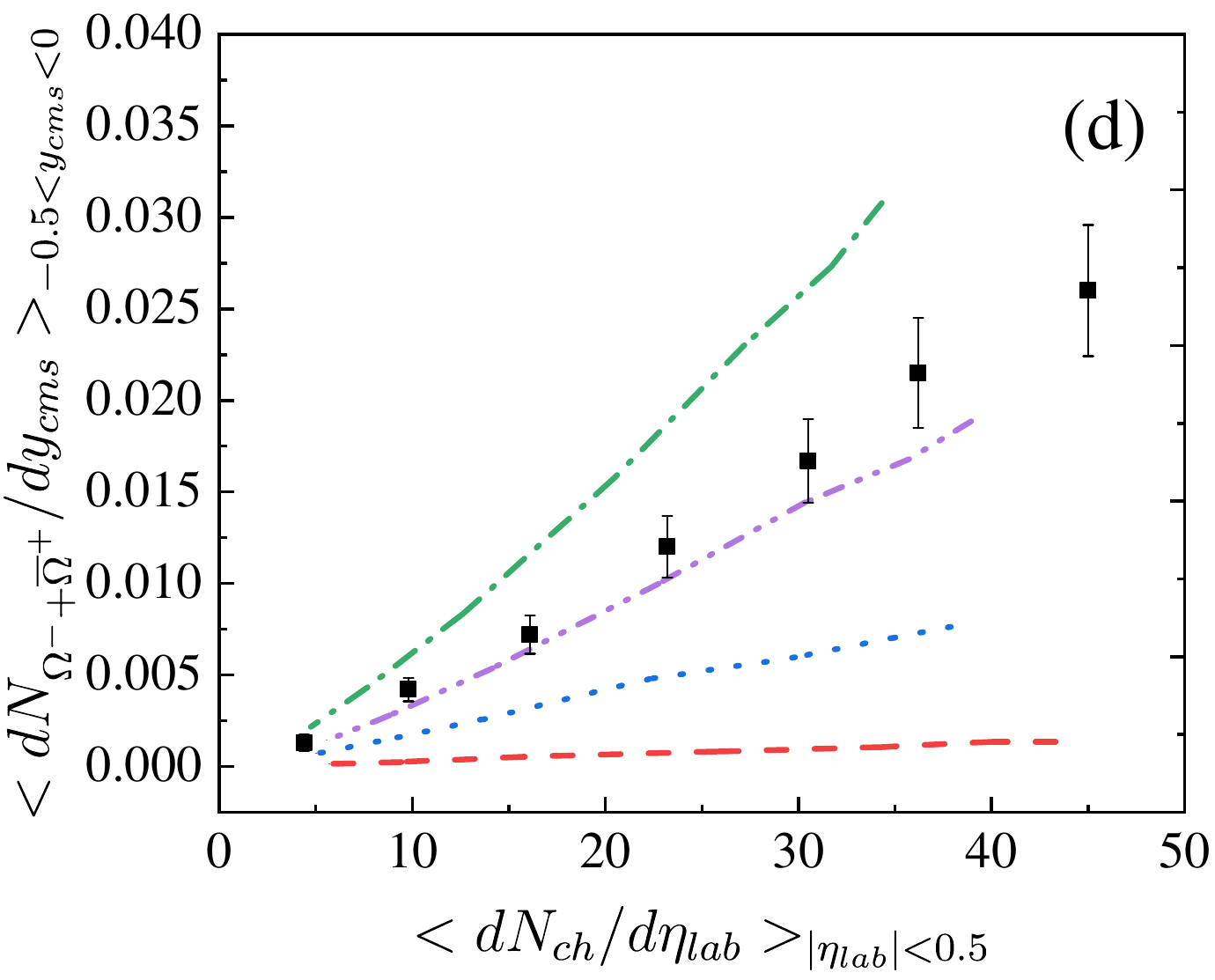}
		\label{fig2:d}
	}
        
	\caption{(Color online) Multiplicity dependent $ \langle dN/dy_{cms} \rangle $ of $K$(a), $\Lambda$(b), $\Xi$(c), and $\Omega$(d), varying with the $\langle dN_{ch}/d\eta_{lab} \rangle_{|\eta_{lab}|<0.5}$ in NSD $p$Pb collisions at $\sqrt{s_{NN}}=5.02$ TeV.  Four different CR and RH schemes are shown:
MPI-CR, red dashed line; QCD-CR, blue dotted line; QCD-CR+FR, green dash-dotted line; QCD-CR+FR+SS, purple dash-dot-dotted line. The experimental data are taken from Refs.~\cite{ALICE:2013wgn,ALICE:2015mpp}.}
	\label{fig:PACIAE-lambda+xi+omega}
\end{figure*}


%



\subsection{Model Setup}

In consistent with ALICE experiment, the simulations of asymmetric $p$Pb collisions are performed in the laboratory 
frame (lab) with the beam energy of 4 TeV for protons and 1.58 TeV per nucleon for Pb 
nuclei~\cite{ALICE:2013wgn,ALICE:2015mpp, ALICE:2014xrc}, resulting in the 
collision energy of $ \sqrt{s_{NN}} = 5.02 $ TeV in the $NN$ center-of-mass system (cms). The simulated events are chosen as the non-single diffractive (NSD) ones~\cite{ALICE:2013wgn,ALICE:2015mpp}. The tuned parameters used in this work are listed in Table~\ref{tab2}, while the other parameters are set as the default values. The meaning of these parameters is as follows:
\begin{itemize}
  \item MultipartonInteractions: Kfactor is a multiplicative factor of the 
parton-parton scattering cross sections in MPI.
  \item MultiPartonInteractions: pT0Ref is the reference scale of the minimum 
$p_T$ cutoff related to the regularization of the divergence of the QCD cross section for $p_T \rightarrow 0$.
  \item StringPT: sigma corresponds to the width in the Gaussian $p_x$ and 
$p_y$ transverse momentum distributions for primary hadrons in the string 
fragmentation.
  \item StringFlav: probQQtoQ is the suppression of diquark-antidiquark pair production compared to the quark-antiquark production in the string 
fragmentation.
  \item StringFlav: probStoUD is the suppression of $s$ quark pair production compared to the $u$ or $d$ pair production in the string fragmentation.
  \item StringFlav: probSQtoQQ is the extra suppression of strange diquark 
production compared to the normal suppression of strange quark in the string fragmentation.
\end{itemize}

\begin{table}[ht]
    \centering
    \caption{Values of various parameters in $p$Pb collisions at $\sqrt{s_{NN}} = 5.02$ TeV for PACIAE 4.0 model simulations.
    \label{tab2}}
    \setlength{\tabcolsep}{21pt}
    \begin{tabular}{ c | c }
        \hline
        \hline
         Parameters & Value  \\
        \hline
        MultipartonInteractions: Kfactor & 1.1  \\
        MultiPartonInteractions: pT0Ref  & 1.3  \\
        StringPT: sigma                   & 0.2  \\
        StringFlav: probQQtoQ            & 0.07 \\
        StringFlav: probStoUD            & 0.3 \\
        StringFlav: probSQtoQQ           & 0.18  \\
        \hline
        \hline
    \end{tabular}
\end{table}

\begin{figure*}[hbt!]
	\centering

        \subfigure{
		\includegraphics[width=0.45\textwidth]{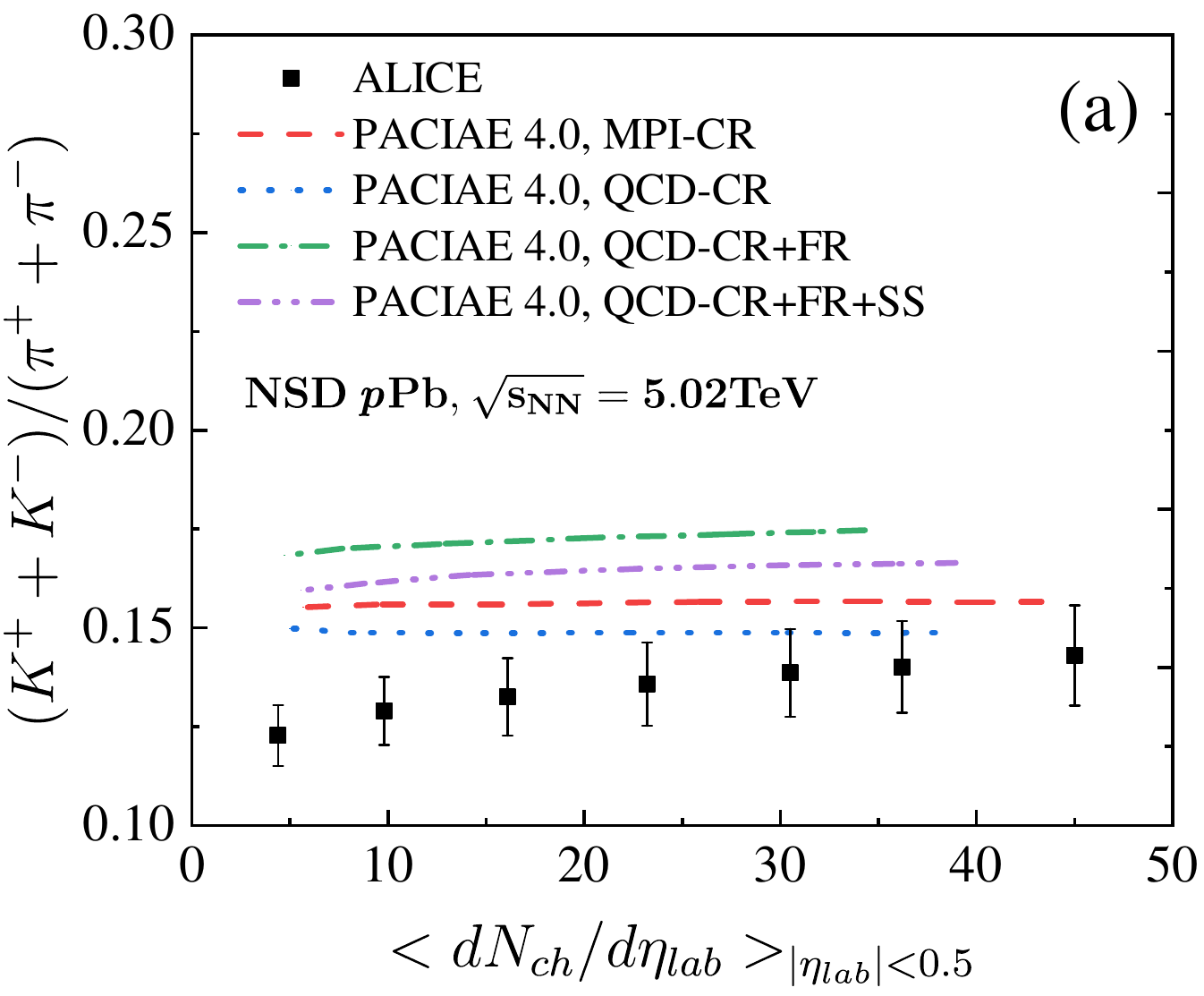}
		\label{fig3:a}
        }  
        \subfigure{
		\includegraphics[width=0.45\textwidth]{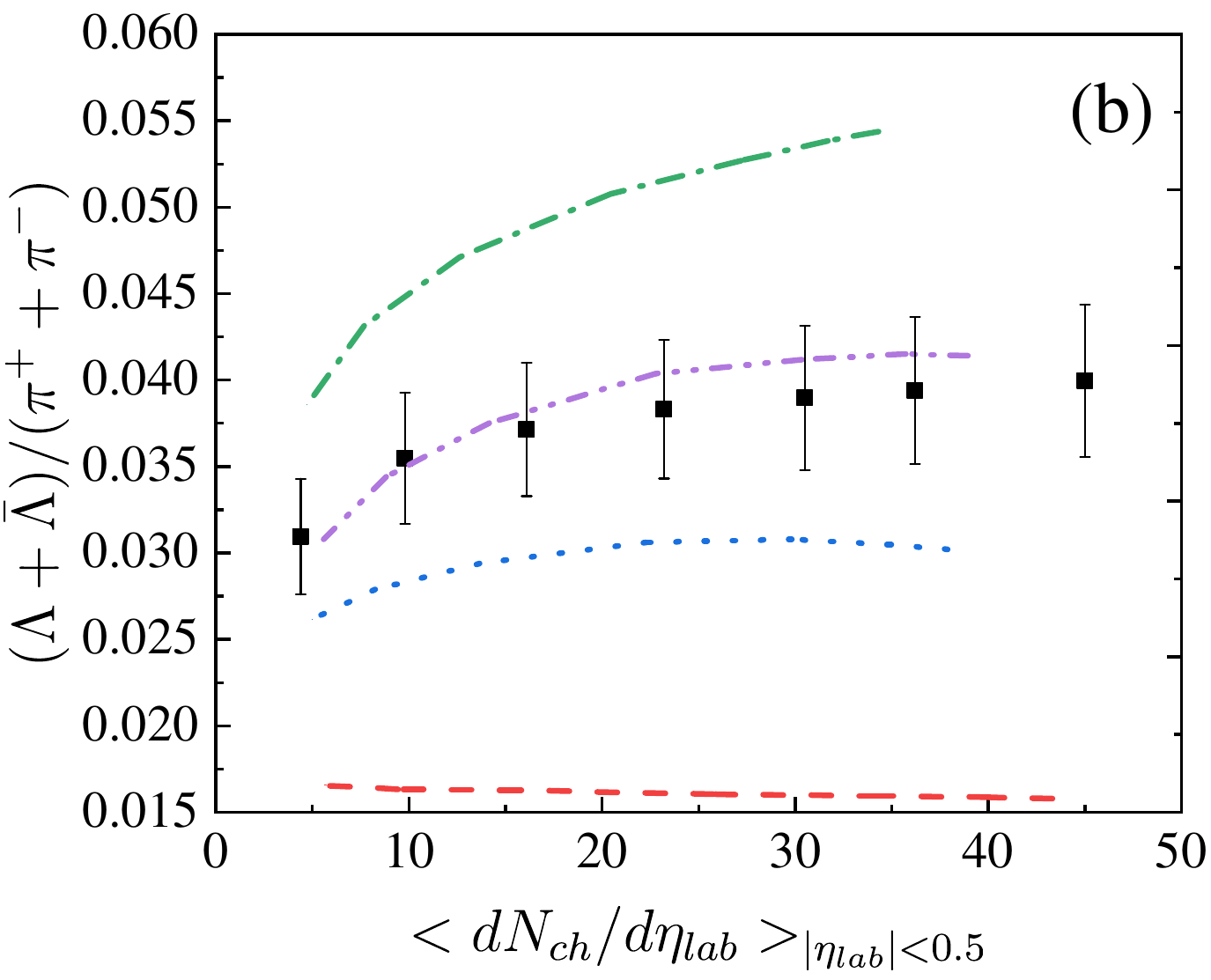}
		\label{fig3:b}
	}
        \subfigure{	\includegraphics[width=0.45\textwidth]{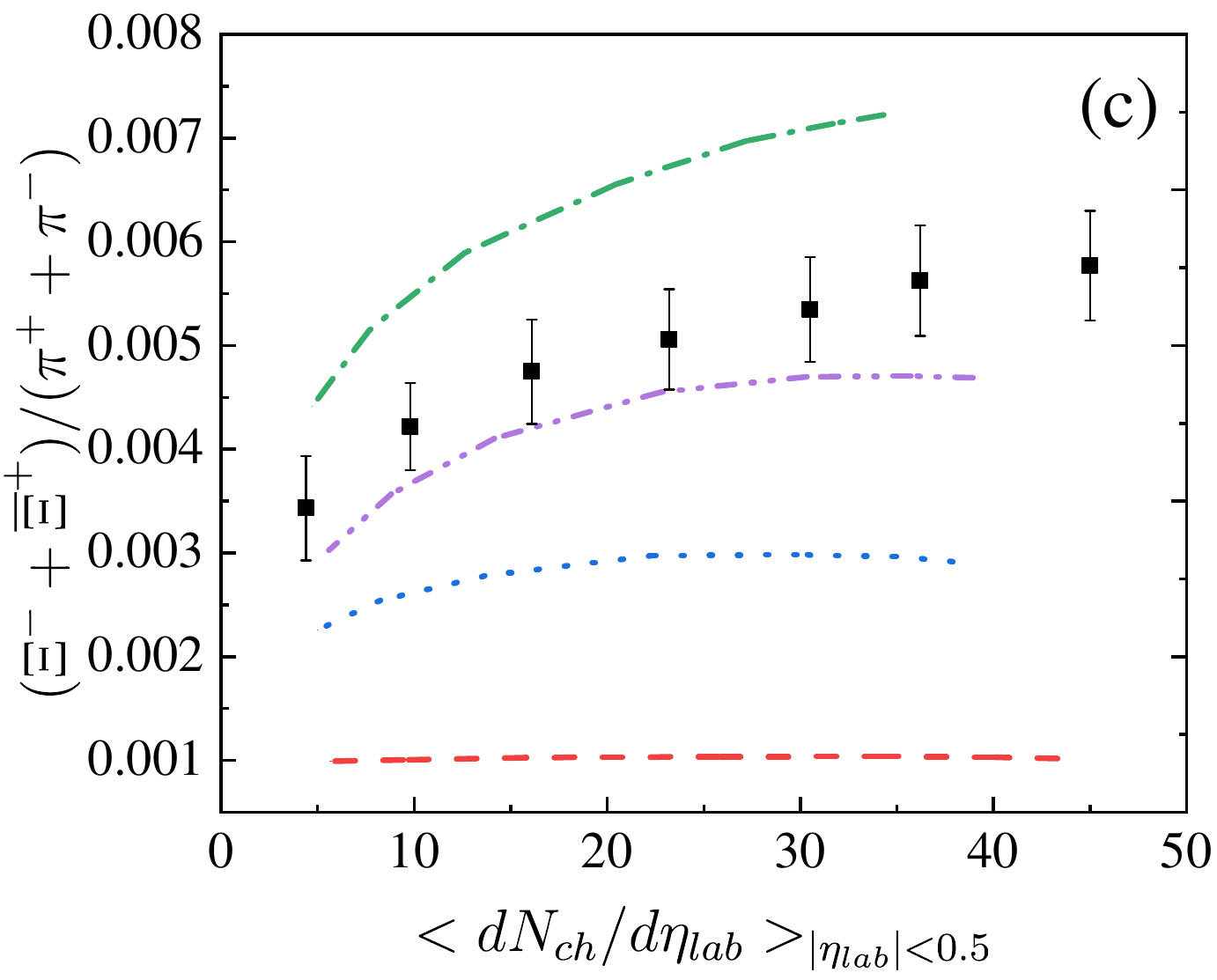}
		\label{fig3:c}
        }
        \subfigure{	\includegraphics[width=0.45\textwidth]{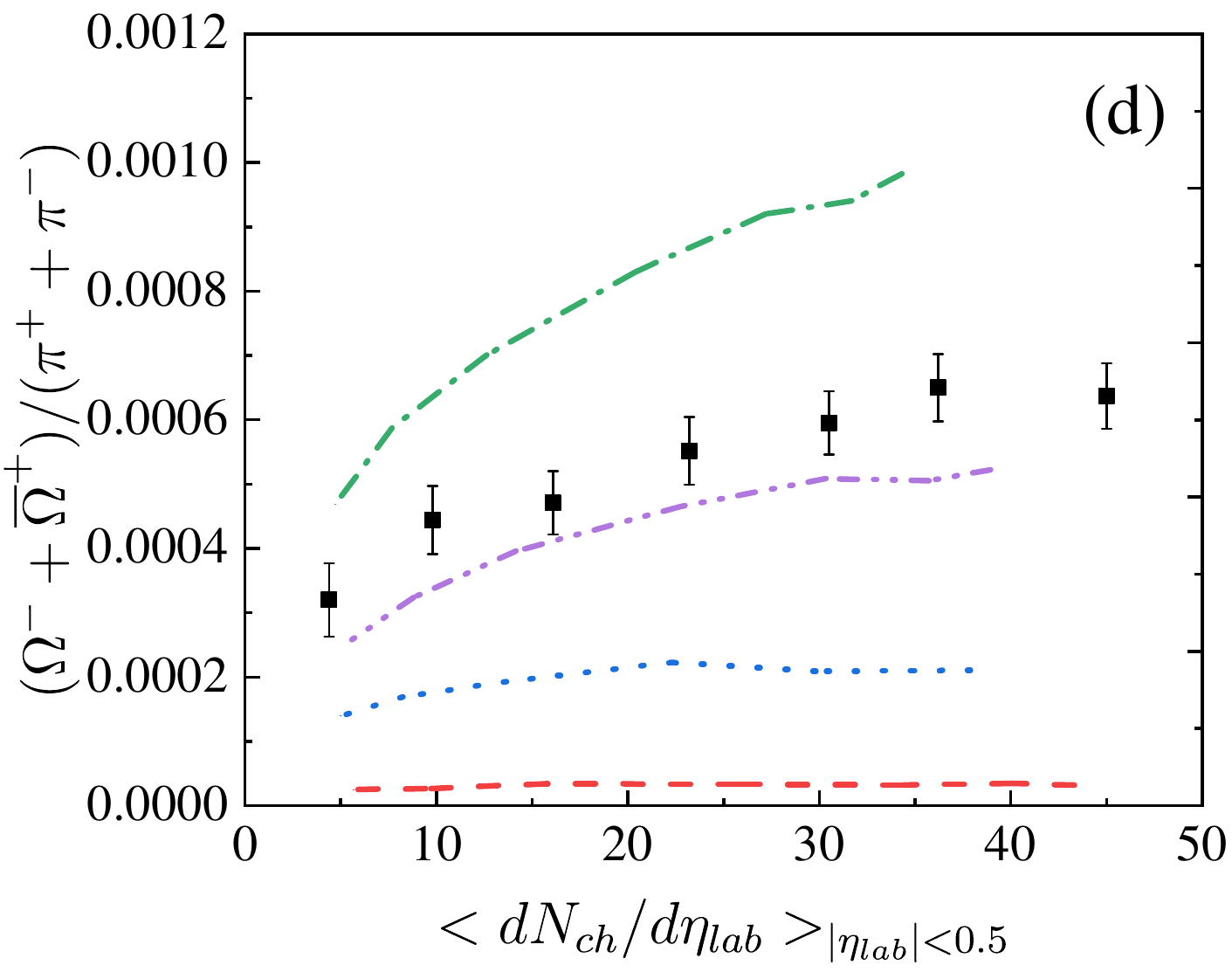}
		\label{fig3:d}
	}
        
	\caption{(Color online) The ratio of yields of strange hadron ($K$(a), $\Lambda$(b), $\Xi$(c) and $\Omega$(d)) -to-pion as a function of $ \langle dN_{ch}/d\eta_{lab} \rangle_{|\eta_{lab}|<0.5}$ in NSD $p$Pb collisions at $\sqrt{s_{NN}}=$ 5.02 TeV using PACIAE 4.0 model. Four different CR and RH schemes are shown:
MPI-CR, red dashed line; QCD-CR, blue dotted line; QCD-CR+FR, green dash-dotted line; QCD-CR+FR+SS, purple dash-dot-dotted line. The experimental data are taken from Refs.~\cite{ALICE:2013wgn,ALICE:2015mpp}.}
	\label{fig:lambda+xi+omega+over+pi}
\end{figure*}

\section{Results AND DISCUSSIONS}
\label{sec:results}

In this work, the data analysis of identified hadrons is performed in the 
center-of-mass system. The rapidity interval for $\pi$, $K$, $\Lambda$ is 
$ 0 < y_{cms} < 0.5$~\cite{ALICE:2013wgn}, while that of $\Xi$ and $\Omega$ is 
$ -0.5 < y_{cms} < 0$~\cite{ALICE:2015mpp}. With the tuned parameters, 
PACIAE 4.0 gives the mean multiplicity of pions of 
$ \langle dN_\pi/dy \rangle = 41.81 $ for 
$0-5\%$ $p$Pb collisions at $\sqrt{s_{NN}} = 5.02$ TeV, consistent with the corresponding ALICE data of $40.81\pm2.26$~\cite{ALICE:2013wgn}.  
The multiplicity dependence of $\pi$ yields versus the mean multiplicity densities of charged 
particles ($\langle dN_{ch}/d\eta \rangle$) within $ | \eta_{lab} | < 0.5 $ is presented in 
Fig. \ref{fig:PACIAE-pi+ks0+p}. The results of $\pi$ from PACIAE 4.0 with four schemes of MPI-CR, QCD-CR, QCD-CR+FR and QCD-CR+FR+SS are shown as the red dashed, blue dotted, green dash-dotted and purple dash-dot-dotted lines, respectively. The ALICE data are shown by black squares with error 
bars~\cite{ALICE:2013wgn}. The $\pi$ yields from all four scenarios show 
reasonable agreement with experimental data, as expected. It demonstrates that the set parameters are rational.

We then analyze the dependence of $K$, $\Lambda$, $\Xi$ and $\Omega$ yields on the mean multiplicity densities of charged particles ($\langle dN_{ch}/d\eta \rangle$) within $| \eta_{lab} | < 0.5$ and compare those with the ALICE experimental data. As shown in 
Fig.~\ref{fig2:a}, for the $ \langle dN/dy \rangle$ distribution of kaons, four mechanisms give rather similar results. For the strange baryons, we see from Fig.~\ref{fig2:b} to ~\ref{fig2:d} that: the MPI-CR scheme systematically underestimates particle production. The QCD-CR scheme enables baryon production through the junction structure formation, where three color flux tubes recombine, leading to enhanced baryon production. 
However, the results show that the QCD-CR scheme still underestimates the yield of
strange baryons. The RH mechanism, when coupled with the QCD-CR, enhances the string tension, consequently increasing strange baryon production yields across all multiplicity classes. Our analysis shows that the QCD-CR+FR scheme overestimates strange baryons production. The combined QCD-CR+FR+SS scheme can provide slightly smaller string tension compared to the QCD-CR+FR scheme. The experimental results for the $\Lambda$ baryon (single strange quark) lie between those predicted by the QCD-CR and QCD-CR+FR+SS mechanisms, whereas those for $\Xi$ and $\Omega$ fall between the QCD-CR+FR+SS and QCD-CR+FR mechanisms. This indicates that heavier strange baryons require a stronger string tension for their production. We emphasize that the experimental data of $K$, $\Lambda$, $\Xi$ 
and $\Omega$ are reproduced reasonably well with the QCD-CR+FR+SS mechanism, simultaneously.

\begin{figure*}[hbt!]
	\centering

        \subfigure{
		\includegraphics[width=0.45\textwidth]{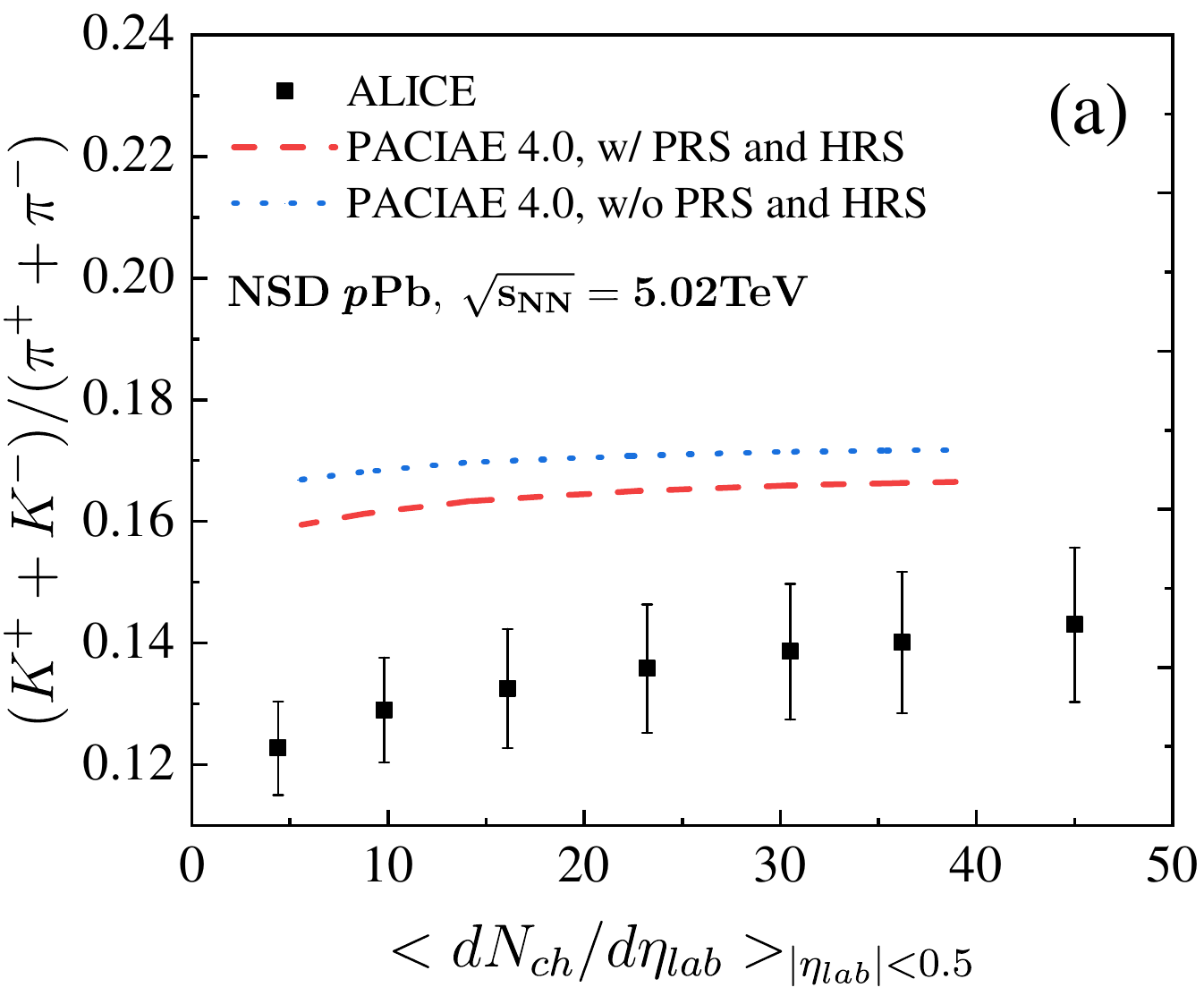}
		\label{fig4:a}
	}
        \subfigure{
		\includegraphics[width=0.45\textwidth]{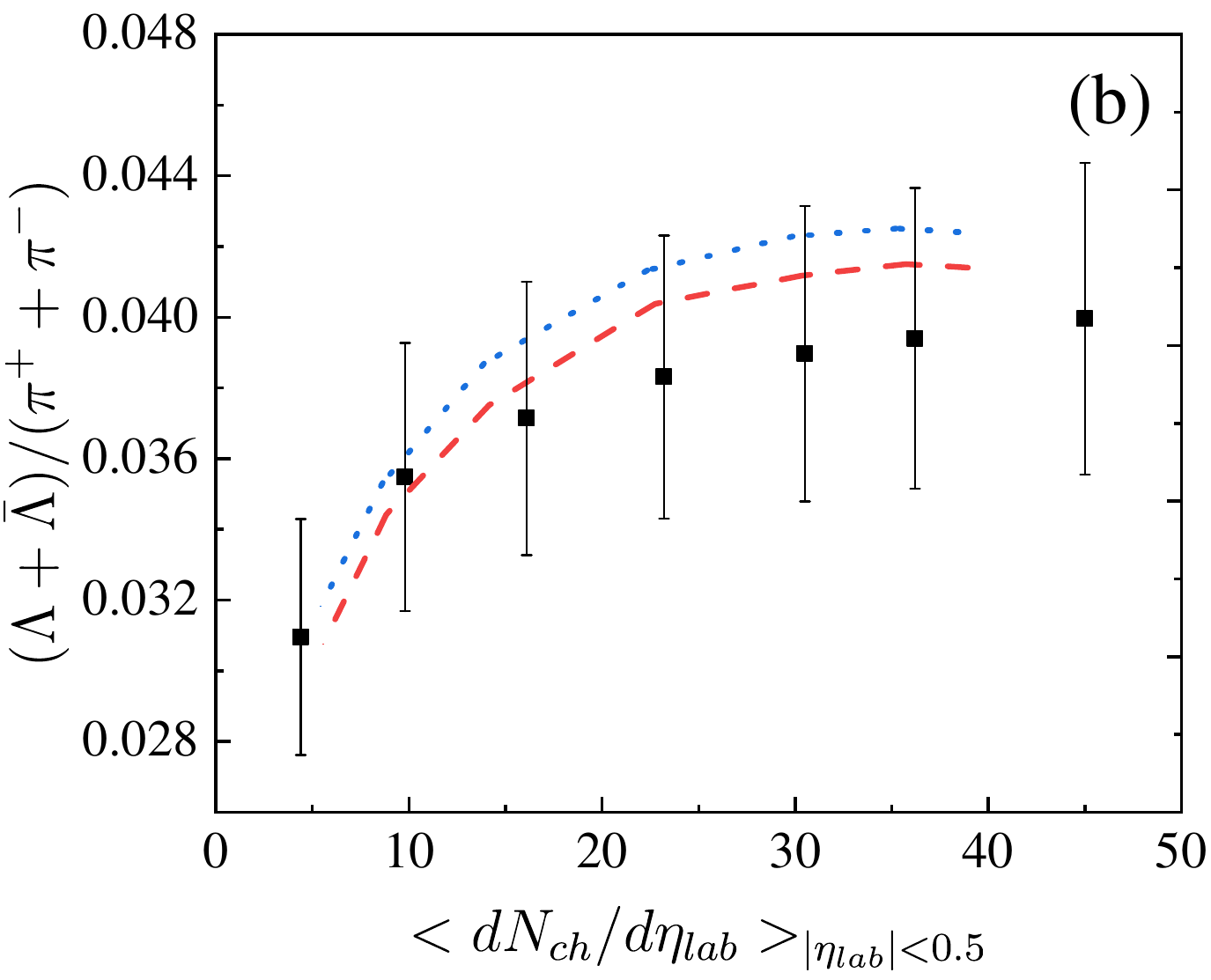}
		\label{fig4:b}
	}
        \subfigure{
		\includegraphics[width=0.45\textwidth]{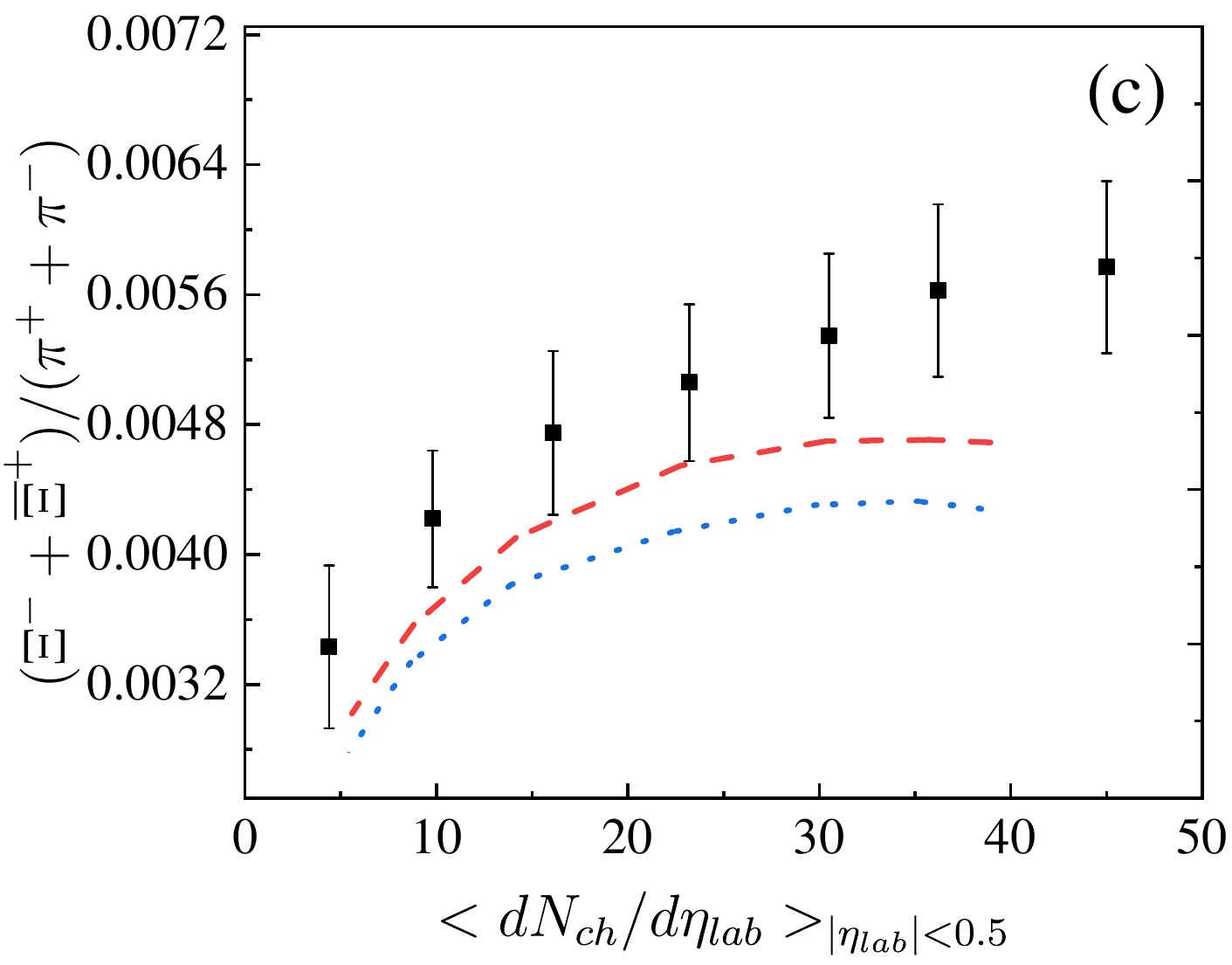}
		\label{fig4:c}
        }
        \subfigure{
	\includegraphics[width=0.45\textwidth]{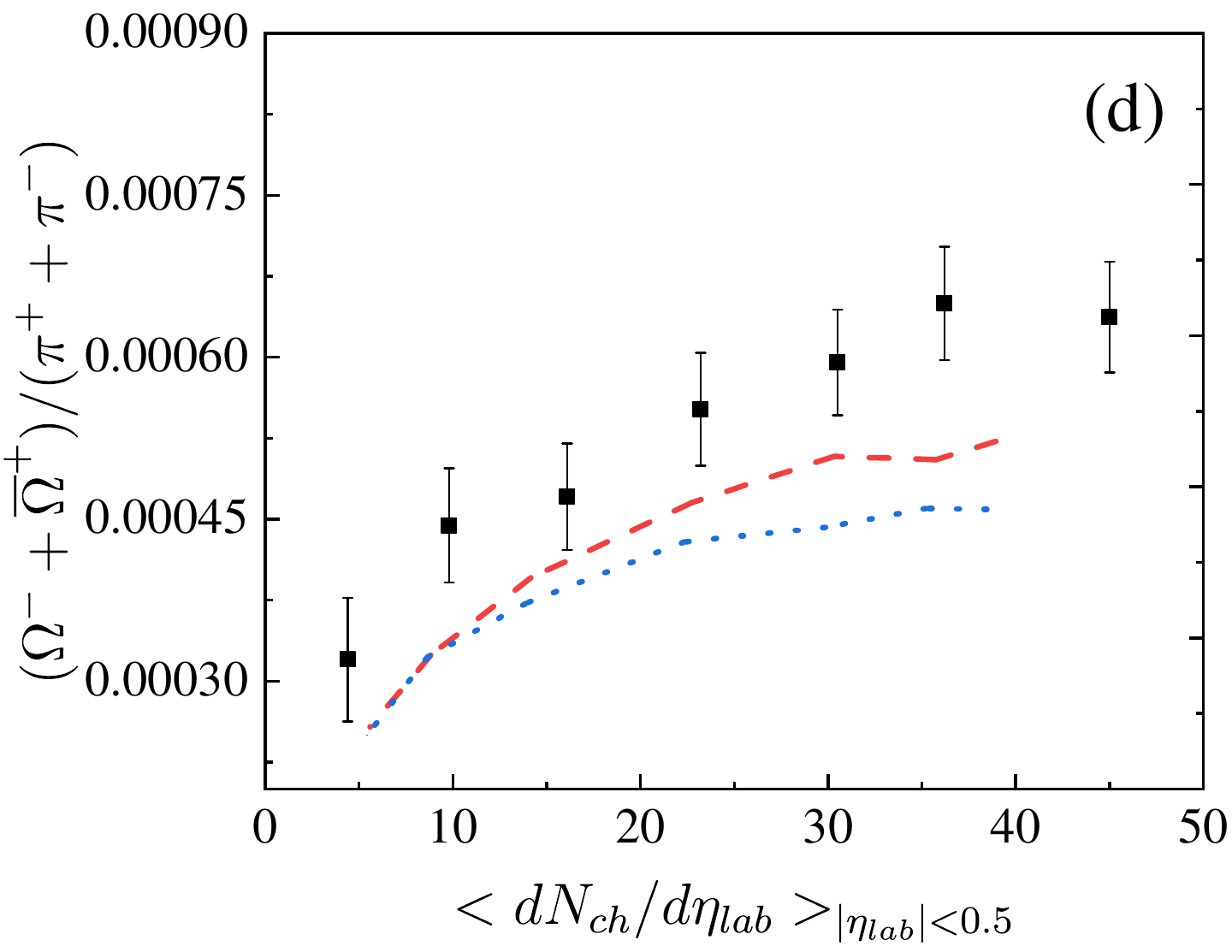}
		\label{fig4:d}
	}
        
	\caption{(Color online) The ratio of yields of strange hadron($K$(a), $\Lambda$(b), $\Xi$(c) and $\Omega$(d)) -to-pion as a function of $<dN_{ch}/d\eta_{lab}>_{|\eta_{lab}|<0.5}$ in NSD $p$Pb collisions at $\sqrt{s_{NN}}=$ 5.02 TeV using PACIAE 4.0 with (w/) (red dashed line) and without (w/o) (blue dotted line) both PRS and HRS processes. The QCD-CR+FR+SS scheme is chosen. The experimental data are taken from Refs.~\cite{ALICE:2013wgn,ALICE:2015mpp}.}
	\label{fig:PYTHIAoverpi}
\end{figure*}

To investigate the strangeness enhancement observed by ALICE in $p$Pb systems, we estimate the yield ratios of strange hadron ($K$, $\Lambda$, $\Xi$ and $\Omega$) to charged $\pi$. In Fig.~\ref{fig:lambda+xi+omega+over+pi}, these 
ratios are plotted as a function of 
$\langle dN_{ch}/d\eta_{lab} \rangle_{|\eta_{lab}|<0.5}$ with different 
simulation schemes and compared to the ALICE 
data~\cite{ALICE:2013wgn,ALICE:2015mpp}. It shows that the MPI-CR scheme 
systematically underestimates the ALICE data, except $K$/$\pi$. This is because the MPI-CR scheme is not expected to generate sufficient baryons. Although reproducing the $K$/$\pi$ ratios in higher multiplicities, the QCD-CR scheme does not describe the ALICE strange baryons enhancement data. The results of the QCD-CR+FR scheme overestimate the ALICE ratios data for all strange baryon-to-pion. The ALICE data of all strange baryons can be well described by the QCD-CR+FR+SS scheme simulations. However, as the number of strange quarks increases (from $\Lambda$ to $\Omega$), the simulated results of the QCD-CR + FR + SS scheme become increasingly lower than the ALICE data. It is consistent with the strange hadron yields in Fig. \ref{fig:PACIAE-lambda+xi+omega}.

We also study the effects of partonic and hadronic rescatterings (PRS and HRS), same as Ref. ~\cite{Xie:2025vnh},  on the strange hadron-to-pion ratios in PACIAE 4.0 with the QCD-CR+FR+SS scheme. There are four combinations with (w/) and/or without (w/o) both the PRS and HRS. For clarity, we show only the results of w/o both PRS and HRS and w/ both PRS and HRS, respectively, in Fig.~\ref{fig:PYTHIAoverpi}. We found that the baryon-to-pion ratios of $\Xi$ and $\Omega$ from simulations w/ PRS and HRS are systematically higher than those w/o PRS and HRS, in contrast to the $K / \pi$ and $\Lambda / \pi$ ratios. It is mainly attributed to HRS, because PRS in this work is limited to elastic processes that do not significantly affect the flavor content of the final partonic state. In HRS, there exist inelastic processes involving strange hadrons~\cite{Sa:2011ye,Lei:2023srp,Lei:2024kam}, e.g. 
$ \pi \Lambda \rightleftharpoons K \Xi $, 
$ K \bar{ \Lambda } \rightleftharpoons \pi \overline{\Xi}$, etc. At the same time, the direct production of single-strange hadrons $K$ and 
$\Lambda$ from hadronization is greater than that of multi-strange baryons 
$\Xi$ and $\Omega$. Part of $ K $ and $ \Lambda $ could be converted to 
$\Xi$ and $\Omega$ through the HRS.

\section{Conclusions}
\label{sec:conclusions}
In this work, we employ the PACIAE 4.0 model with different selected physical mechanisms to investigate the multiplicity dependence of the strange hadron $\langle dN/dy \rangle$ as well as the ratios of strange hadrons to charged pions, and compare them with the corresponding ALICE data. The multiplicity-dependent $\langle dN/dy \rangle$ of kaons, four mechanisms give rather similar results. MPI-CR mechanism underestimates the yields of strange baryons ($\Lambda$, $\Xi$ and $\Omega$). The QCD-CR scheme incorporating the junction topology gives an enhanced baryon production, but it is still not enough to describe the integrated yields of strange baryons. Furthermore, we found the combination of QCD-CR and the rope hadronization (RH) regime including the flavor ropes (FR) and the string shoving (SS) (QCD-CR+FR+SS) provides a good description of the production of strange baryons. The results of the ratios of strange baryon-to-pion also confirm the same conclusion as those for the yields. We have also investigated the role of partonic and hadronic rescatterings. The results indicate that the effects of the PRS on the production of multi-strange baryons can be neglected. However, the HRS exerts a promoting effect on the production of multi-strange baryons ($\Xi$ and $\Omega$) due to the inelastic processes involved.

Our results demonstrate that the color-reconnection, rope hadronization, and two rescattering processes implemented in the PACIAE 4.0 model successfully reproduce the characteristic enhancement of strange particle production. These implementations provide novel approaches to investigate the phenomena of strangeness enhancement in the $p$A small collision system and fresh insights into its underlying mechanisms. The similar study is worth extending to the Pb-Pb collisions at LHC energies in the future.

\begin{acknowledgments}
This work is supported by the National Natural Science Foundation of China (Grant No. 12375135). W. Lei
 would like to thank Dr. Jia-Li Deng for valuable discussions. Y.-L. Yan acknowledges funding from the Continuous Basic Scientific Research Project (Grant No. WDJC-2019-13). H.Zheng acknowledges financial support from the Key Laboratory of Quark and Lepton Physics at Central China Normal University (Grant No. QLPL2024P01). W.-C. Zhang is supported by the Natural Science Basic Research Program of Shaanxi Province, China (Program No. 2023-JCYB-012). 
\end{acknowledgments}


\end{document}